\documentclass[preprint,prd]{revtex4}%
\usepackage{amsmath}
\usepackage{amsfonts}
\usepackage{amssymb}
\usepackage{graphicx}
\def\be{\begin{equation}}
\def\ee{\end{equation}}
\def\bea{\begin{eqnarray}}
\def\eea{\end{eqnarray}}

\begin{document}

\title{Fundamentals of Geometrothermodynamics}

\author{Hernando  Quevedo$^{1,2}$ and Mar\'\i a N. Quevedo$^3$}
\email{quevedo@nucleares.unam.mx,  maria.quevedo@unimilitar.edu.co}
\affiliation{
$^1$Instituto de Ciencias
Nucleares, Universidad Nacional Aut\'onoma de M\'exico,
 AP 70543, M\'exico, DF 04510, Mexico\\
$^2$ICRANet, Dipartimento di Fisica, Universit\`a di Roma "La
Sapienza",  I-00185 Roma, Italy\\
$^3$Departamento de Matem\'aticas \\
Universidad Militar Nueva Granada    \\
Cra. 11 No. 101-80, Bogot\'a D.E., Colombia}

\date{\today}

\begin{abstract}
We present the basic mathematical elements of geometrothermodynamics which is 
a formalism developed to describe in an invariant way the thermodynamic 
properties of a given thermodynamic system in terms of geometric structures.
First,  in order to represent the first law of thermodynamics and the general Legendre 
transformations in an invariant way, we define the phase manifold 
as a Legendre invariant Riemannian manifold with a contact structure. 
The equilibrium manifold is defined by using a harmonic map which includes
the specification of the fundamental equation of the thermodynamic system.
Quasi-static thermodynamic processes are shown to correspond to geodesics of the
equilibrium manifold which preserve the laws of thermodynamics.
We study in detail the equilibrium manifold of the ideal gas and the van der Waals 
gas as concrete examples of the application of geometrothermodynamics.

\end{abstract}

\maketitle

%%%%%%%%%%%%%%%%%%%%%%%%%%%%%%%%%%%%%%%%%%%%%%%%%%%%%%%%%%%%%%%%%%%%%%%%%%%%%%%%%%%%%%%%%%%%%%
%%%%%%%%%%%%%%%%%%%%%%%%%%%%%%%%%%%%%%%%%%%%%%%%%%%%%%%%%%%%%%%%%%%%%%%%%%%%%%%%%%%%%%%%%%%%%%

\section{Introduction}
\label{sec:int}

Differential geometry is a very important tool of mathematical physics with 
many applications in physics,
chemistry and engineering. As an example, one can mention the case
of the four known interactions of nature which can be described in terms of geometrical concepts. 
Indeed, Einstein proposed in 1915 the astonishing principle ``field strength = curvature" 
to understand the physics of the gravitational field (see, for instance, Ref. \cite{frankel}). 
In an attempt to associate a geometric structure to  the electromagnetic field, Yang and Mills \cite{ym53}
used in 1953 the concept of a principal fiber bundle with the Minkowski spacetime as the base manifold and the symmetry group $U(1)$
as the standard fiber to demonstrate that the Faraday tensor can be interpreted as the curvature of this particular fiber bundle.
Today, it is well known \cite{frankel} that the weak interaction and the strong interaction can be represented as the 
curvature of a principal fiber bundle with a Minkowski base manifold and the standard fiber $SU(2)$ and $SU(3)$, respectively.
In this work, we will show that it is possible to interpret the thermodynamic interaction as the curvature of 
a Legendre invariant Riemannian manifold. It should be mentioned that our interpretation of the thermodynamic interaction 
is based upon the standard statistical approach 
to thermodynamics in which all  the properties of the system can be derived from the explicit form of the corresponding 
Hamiltonian and partition function \cite{greiner},
and in which the interaction between the particles of the system is described by the potential part of the Hamiltonian. 
Consequently, if the potential vanishes, we say that the system has a zero thermodynamic interaction.

In very broad terms, one can say that in a thermodynamic system, all the known forces act among the particles that constitute the system.  
Due to the large number of particles involved in the system, only a statistical approach is possible, from which average values for
the physical quantities of interest are derived. 
Although the laws of thermodynamics are  based entirely upon empirical results which are satisfied under certain
conditions in almost any macroscopic system, the geometric approach to thermodynamics
has proved to be very useful. One can say that the following three branches of geometry
have found sound applications
in equilibrium thermodynamics: analytic geometry, Riemannian geometry, and contact geometry.

 Probably, one of the most important contributions of analytic geometry
to the understanding of thermodynamics is the identification of points of phase transitions with
extremal points of the surface determined by the corresponding state equation. 
For a more detailed description of these contributions see, for instance, \cite{callen,huang}.
Riemannian geometry was first introduced in statistical physics and thermodynamics
by Rao \cite{rao45}, in 1945, by means of a metric whose components in local coordinates coincide with Fisher's
information matrix. Rao's original work has been followed up and extended by a number of
authors (see, e.g., \cite{amari85} for a review). On the other hand, Riemannian geometry in the space of
equilibrium states was introduced by Weinhold \cite{wei75} and Ruppeiner \cite{rup79,rup95},
who defined metric structures as the Hessian of the internal energy and the entropy, respectively.
Both metrics have been used intensively to study the geometry of the thermodynamics of
ordinary systems and black holes; however, several inconsistencies and contradictions have been found 
\cite{am,ama,aman06a,scws,caicho99,sst,med,mz,hernando2}. It is now well 
established that these puzzling results are a consequence of the fact that Weinhold and Ruppeiner metrics 
are not invariant with respect to Legendre transformations \cite{quev07}. 
Furthermore, contact geometry was introduced by Hermann \cite{her73} into the thermodynamic phase space
in order to formulate in a consistent manner the geometric version of the laws of thermodynamics.

In order to incorporate Legendre invariance in Riemannian structures
at the level of the phase space and the equilibrium space, the formalism of
geometrothermodynamics (GTD)
 was recently proposed by Quevedo \cite{quev07}. The main motivation for
introducing the formalism of GTD was to formulate a geometric approach which takes into account
the fact that in ordinary thermodynamics the description of a system does not depend
on the choice of the thermodynamic potential, i. e., it is invariant with respect to Legendre 
transformations. 
 One of the main goals of GTD has been to interpret in an
invariant manner the curvature of the equilibrium space as a manifestation of the thermodynamic
interaction. This would imply that an ideal gas and its generalizations with no mechanic interaction
correspond to a Riemannian manifold with vanishing curvature. Moreover, in the case of interacting
systems with non-trivial structure of phase transitions, the curvature should be non-vanishing and
reproduce the behavior near the points where phase transitions occur. These intuitive statements
represent concrete mathematical conditions for the metric structures of the phase and equilibrium spaces. 
In the present work, we present geometric structures which satisfy these conditions for
systems with no thermodynamic interaction as well as for systems characterized by
interaction with first and second order phase transitions.

In this work, we present the formalism of GTD by using  Riemannian contact geometry for the definition of the thermodynamical phase 
manifold and harmonic maps for the definition of the equilibrium manifold. We will see that this approach allows us 
to interpret any thermodynamic system  as a hypersurface in the equilibrium space completely determined by the
field theoretical approach of harmonic maps. 
This paper is organized as follows: In Section \ref{sec:rcg}, we introduce the main concepts of Riemannian contact geometry 
that are necessary to define the phase manifold. Section \ref{sec:eqm} is dedicated to the description of 
the equilibrium manifold as resulting from a harmonic map in which the target space is the phase manifold.
Section \ref{sec:qua} contains a discussion of the quasi-static thermodynamic processes which are
interpreted as geodesics preserving the laws of thermodynamics. 
In Section \ref{sec:gases}, we present the main geometric properties of the ideal and the van der Waals gas. 
Finally, Section \ref{sec:con} is devoted
to discussions of our results and suggestions for further research.
Throughout this paper, we use units in which $G=c=k_{_B}=\hbar =1$.

%%%%%%%%%%%%%%%%%%%%%%%%%%%%%%%%%%%%%%%%%%%%%%%%%%%%%%%%%%%%%%%%%%%%%%%%%%%%%%%%%%%%%%%%
%%%%%%%%%%%%%%%%%%%%%%%%%%%%%%%%%%%%%%%%%%%%%%%%%%%%%%%%%%%%%%%%%%%%%%%%%%%%%%%%%%%%%%%%
\section{The thermodynamic phase manifold} 
\label{sec:rcg}

Consider a $(2n+1)-$dimensional differential manifold ${\cal T}$ and its tangent manifold $T({\cal T})$. 
Let ${\cal V}\subset T({\cal T})$ be an arbitrary  field of hyperplanes on ${\cal T}$. 
It can be shown that there exists a non-vanishing differential 1-form $\Theta$ on the cotangent manifold 
$T^*({\cal T})$ such that the field ${\cal V}$ can be associated with the kernel of $\Theta$, i. e.,
${\cal V} = \ker \Theta$. 
If the Frobenius integrability condition $\Theta\wedge d \Theta =0$ is satisfied, the hyperplane field ${\cal V}$ is
completely integrable. On the contrary, if   $\Theta \wedge d \Theta \neq 0$, then ${\cal V}$ is non-integrable. 
In the limiting case $\Theta \wedge (d \Theta)^n \neq 0$, the hyperplane field ${\cal V}$ becomes maximally non-integrable
and  is said to define a contact structure on ${\cal T}$. 
The pair $({\cal T},{\cal V})$ determines a contact manifold \cite{handbook}
and is sometimes denoted as $({\cal T},\Theta)$
to emphasize the role of the contact form $\Theta$. 
Consider $G$ as a non-degenerate metric on ${\cal T}$. 
The set $({\cal T}, \Theta, G)$ defines a Riemannian contact manifold.
It should be noted that the condition
$\Theta \wedge (d\Theta)^n \neq 0$ is independent of $\Theta$; in fact, it is a property of ${\cal V}=\ker\Theta$.
If another 1-form $\Theta'$ generates the same ${\cal V}$, it must be of the form 
$\Theta' = f \Theta$, where $f: {\cal T} \rightarrow \mathbb{R}$ is a smooth non-vanishing function. This implies 
that the contact manifold $({\cal T}, \Theta)$ is  uniquely defined up to a smooth function 
$f: {\cal T} \rightarrow \mathbb{R}$.

Let us choose a particular set of  coordinates of ${\cal T}$ as $Z^A =\{\Phi, E^a, I ^a\}$ with $a=1,...,n$, and $A=0,1,...,2n$. 
Here, $\Phi$ represents the thermodynamic potential used to describe the system
whereas the coordinates $E^a$ correspond to the extensive variables and $I^a$ to the
intensive variables.  Notice that since in the phase manifold ${\cal T}$ all the coordinates $\Phi$, $E^a$ and $I^a$ 
must be completely independent, it is not possible to describe thermodynamic systems  in ${\cal T}$ which are usually 
defined in terms of equations of state that relate different thermodynamic variables. An important ingredient of GTD
is the concept of Legendre transformations that in general are defined as \cite{arnold}
\be
\{Z^A\}\longrightarrow \{\widetilde{Z}^A\}=\{\tilde \Phi, \tilde E ^a, \tilde I ^ a\}\ ,
\ee
\be
 \Phi = \tilde \Phi - \delta_{kl} \tilde E ^k \tilde I ^l \ ,\quad
 E^i = - \tilde I ^ {i}, \ \  
E^j = \tilde E ^j,\quad   
 I^{i} = \tilde E ^ i , \ \
 I^j = \tilde I ^j \ ,
 \label{leg}
\ee
where $i\cup j$ is any disjoint decomposition of the set of indices $\{1,...,n\}$,
and $k,l= 1,...,i$. In particular, for $i=\{1,...,n\}$ and $i=\emptyset$, we obtain
the total Legendre transformation and the identity, respectively. 

In the particular coordinates $Z^A=\{\Phi, E^a, I ^a\}$, the contact 1--form can be written as 
\be
 \Theta= d\Phi - \delta_{ab} I^a d E^b \ ,\quad \delta_{ab}={\rm diag} (1,1,...,1)\ ,
\label{gibbs}
\ee
where we assume the convention of summation over repeated indices. This
expression for the 1-form $\Theta$ is manifestly invariant with respect to the Legendre 
transformations given in Eq.(\ref{leg}), i. e., under a Legendre transformation it transforms as
$\Theta\rightarrow \tilde\Theta =  d\tilde \Phi - \delta_{ab} \tilde I^a d \tilde E^b$.  
Consequently, the contact manifold $({\cal T}, \Theta)$ is a Legendre
invariant structure. Furthermore, if we demand the 
Legendre invariance of the metric $G$, the Riemannian contact manifold $({\cal T}, \Theta,G)$ is 
Legendre invariant.
Any Riemannian contact manifold $({\cal T}, \Theta,G)$ whose components are Legendre invariant is 
called a thermodynamic {\it phase manifold} and constitutes the starting point for a description 
of thermodynamic systems in terms of geometric concepts. We would like to emphasize the fact that Legendre invariance
is an important condition that guarantees that the description does not depend on the choice of the 
thermodynamic potential, a property that is essential in ordinary thermodynamics.

From the above description if follows that the only freedom in the construction of the 
phase manifold is in the choice of the metric $G$. Although Legendre invariance 
implies a series of algebraic conditions for the metric components $G_{AB}$ 
\cite{quev07}, and it can be shown that these conditions are not trivially satisfied, 
the  metric $G$ cannot be fixed uniquely. It is important  to mention that
a straightforward computation shows that the flat 
metric $G=\delta_{AB}dZ^A dZ^B$ is not invariant with respect to the Legendre transformations given 
in Eq.(\ref{leg}).  It then follows that the phase manifold is necessarily curved. 
We performed a detailed analysis
of the Legendre invariance conditions and found as a solution the metric 
\be
\label{ginv2}
G=\left(d\Phi - I_a dE^a\right)^2 + \Lambda\, (E_a I_a)^{2k+1} d E^a d I^a \ ,\quad E_a =\delta_{ab}E^b\ ,\quad
I_a =\delta_{ab} I ^b\ ,
\ee
where $\Lambda$ is an arbitrary Legendre invariant real function of $E^a$ and $I^a$, 
and $k$ is an integer.  
To our knowledge, this is the most general metric satisfying the conditions of Legendre invariance. 

If we limit ourselves to the case of total Legendre transformations, 
we find that there exists a class of metrics, 
\begin{equation}
G = \left(d\Phi - I_a dE^a\right)^2  +\Lambda
\left(\xi_{ab}E^{a}I^{b}\right)\left(\chi_{cd}dE^{c}dI^{d}\right) \  
\label{gup1}
\end{equation}
parametrized by the diagonal constant tensors $\xi_{ab}$ and $\chi_{ab}$, which is invariant for several choices 
of these free tensors. In fact, since $\xi_{ab}$ and $\chi_{ab}$ must be constant and diagonal it seems reasonable 
to express them in terms of the usual Euclidean and pseudo-Euclidean metrics 
$\delta_{ab}={\rm diag}(1,...,1)$ and $\eta_{ab} = {\rm diag}(-1, 1, ..., 1)$, respectively. Then, for instance, the choice
\be
\xi_{ab}=\delta_{ab}\ , \quad \chi_{ab}=\delta_{ab}\ 
\label{mfo}
\ee
corresponds to a Legendre invariant metric which has been used to describe the geometric properties of systems with 
first order phase transitions \cite{quev07,qstv10a}. Moreover, the choice
\be
\xi_{ab}=\delta_{ab}\ ,\quad \chi_{ab}=\eta_{ab}\ 
\label{mso}
\ee
turned out to describe correctly second order phase transitions especially in black hole thermodynamics \cite{qstv10a,aqs08,vqs09,qsv09}. 
The additional choice
\be
\xi_{ab}= \frac{1}{2}\left(\delta_{ab}-\eta_{ab}\right)\ ,\quad
\chi_{ab}=\eta_{ab}
\label{msot0}
\ee
can be used to handle in a geometric manner second order phase transitions and also the thermodynamic 
limit $T\rightarrow 0$. Obviously, for a given thermodynamic system it is very important to 
choose the appropriate  metric in order to describe correctly the thermodynamic properties in 
terms of the geometric properties in GTD.

%%%%%%%%%%%%%%%%%%%%%%%%%%%%%%%%%%%%%%%%%%%%%%%%%%%%%%%%%%%%%%%%%%%%%%%%%%%%%%%%%%%%%
%%%%%%%%%%%%%%%%%%%%%%%%%%%%%%%%%%%%%%%%%%%%%%%%%%%%%%%%%%%%%%%%%%%%%%%%%%%%%%%%%%%%%%

\section{The equilibrium manifold}
\label{sec:eqm}

Consider the (smooth) harmonic map 
$\varphi: {\cal E}\rightarrow {\cal T}$, where ${\cal E}$ is a subspace of the phase manifold $({\cal T},\Theta,G)$ 
and $\dim({\cal E}) = n$, where $n$ is the number of independent degrees of freedom of the thermodynamic system, i. e., the 
number of independent thermodynamic variables which are necessary to describe a thermodynamic system. 
Let us assume that the extensive variables $\{E^a\}$ can be used as the coordinates of  the base space ${\cal E}$.
Then, in terms of coordinates, the harmonic embedding map reads $ \varphi :  \{E^a\} \longmapsto \{Z^A(E^a)\}=\{\Phi(E^a), E^a, I^a(E^a)\}$. Since the phase manifold is endowed with a Legendre invariant nondegenerate metric $G$,  
the pullback $\varphi^*$ of the harmonic map induces canonically a thermodynamic metric $g$ on  ${\cal E}$ by means of 
\be 
g=\varphi^* (G)\ ,\quad {\rm i.e.}\quad 
g_{ab} = \frac{\partial Z^A}{\partial E^a} \frac{\partial Z^B}{\partial E^b} G_{AB} 
= Z^A_{,a} Z^B_{,b} G_{AB} \ .
\ee

If we assume that the metric of the base manifold coincides with the induced metric $g=\varphi^*(G)$, 
the action of the harmonic map \cite{misner} can be expressed as  
\be
S = \frac{1}{2}\int d^n E \sqrt{|\det(g)|}\ G_{AB} \frac{\partial Z^A}{\partial E^a}\frac{\partial Z^B}{\partial E^b}
g^{ab}  = \frac{n}{2}\int d^n E \sqrt{|\det(g)|}\ ,
\label{ng}
\ee
and turns out to correspond to the volume element of the submanifold ${\cal E}\subset {\cal T}$. 
Consequently, according to the definition of harmonic maps \cite{misner}, the variation
$\delta S = 0$, i. e., the field equations
\be
\frac{1}{\sqrt{|\det(g)|}}\left(\sqrt{|\det(g)|}\,\, g^{ab}Z^A_{,a}\right)_{,b} + 
\Gamma^A_{\ BC} Z^B_{,b}Z^C_{,c} g^{bc} =0 \ ,
\label{meng}
\ee
represent the condition for ${\cal E}$ to be an extremal hypersurface in the phase manifold 
${\cal T}$ \cite{vqs09}. Here, the symbols $\Gamma^A_{\ BC}$ represent the Christoffel symbols 
associated with the metric $G_{AB}$ of the phase manifold, i. e.,
\be
\Gamma^A_{\ BC} = \frac{1}{2} G^{AD}\left( \frac{\partial G_{DB}}{\partial Z^C}
+\frac{\partial G_{CD}}{\partial Z^B}
-\frac{\partial G_{BC}}{\partial Z^D}\right)\ .
\ee

The pair $({\cal E},g)$ is called equilibrium manifold if the harmonic map $\varphi: {\cal E}\rightarrow {\cal T}$ satisfies the 
condition 
\be 
\varphi^*(\Theta)=\varphi^*(d\Phi -\delta_{ab}\, I^a\, dE^b) = 0\ .
\ee
The last condition implies that 
\begin{equation}
 d\Phi
=I_{a}dE^{a}\ , \quad \frac{\partial \Phi}{\partial E^{a}}=I_{a}\ .
\label{firstlaw}
\end{equation}
The first of these equations corresponds to the first law of
thermodynamics whereas the second one is usually known as the
condition for thermodynamic equilibrium \cite{callen}. 

We see that the harmonic map $\varphi:{\cal E} \rightarrow {\cal T}$ 
defines the equilibrium manifold $({\cal E},g)$ as an extremal submanifold of the phase manifold $({\cal T},\Theta,G)$ in which 
the first law of thermodynamics and the equilibrium conditions hold. This means that the thermodynamic systems are represented through 
the equilibrium manifold and that the phase manifold is an auxiliary geometric structure that allows us to handle correctly the
Legendre transformations and to define the equilibrium manifold in an invariant manner. The harmonic map $\varphi$ 
demands the existence of the function $\Phi=\Phi(E^a)$ that is known in ordinary thermodynamics as the fundamental equation
from which all the equations of state can be obtained \cite{callen}. The second law of thermodynamics implies that the fundamental 
equation satisfies the condition
\be
\pm \frac{\partial^2\Phi}{\partial E^a \partial E^b} \geq 0 \ ,
\ee
where the sign depends on the thermodynamic potential. For instance, if $\Phi$ is identified as the entropy, the sign must be positive whereas it 
is negative if $\Phi$ is the internal energy of the system \cite{callen}.

The metric $g$ of the equilibrium manifold is determined uniquely from the metric $G$ by means of $g=\varphi^*(G)$. Therefore, 
the invariance of $G$ under Legendre transformations implies the invariance of $g$. However, as mentioned above, Legendre transformations
act only on the phase manifold and so to investigate the invariance of $g$ it is necessary to apply Legendre transformations on 
the metric $G$ in ${\cal T}$ that generates $g$. The pullback $\varphi^*$ of the Legendre invariant metric (\ref{ginv2}) generates 
the following thermodynamic metric 
\be
g=\Lambda \left(E_a\Phi_a\right)^{2k+1} \delta^{ab}\Phi_{bc} dE^a dE^c \ ,
\label{gdown1}
\ee
where 
\be 
\Phi_a = \frac{\partial \Phi}{\partial E^a}\ , \quad
\Phi_{bc} = \frac{\partial^2 \Phi}{\partial E^b\partial E^c}\ ,
\ee
which can be shown to be invariant with respect to arbitrary (partial and total) Legendre transformations. On the other hand, 
the metric (\ref{gup1}) of the phase manifold generates the thermodynamic metric
\be
g=\Lambda   \left(\xi^{\ b}_{a} E^a\Phi_b\right)\left(\chi_a^{\ b} \Phi_{bc} dE^a d E^c\right)\ ,
\label{gdown2}
\ee
where 
\be
\xi^{\ b}_{a} =\xi_{ac}\delta^{bc}\ , \quad \chi^{\ b}_{a} =\chi_{ac}\delta^{bc}\ ,
\ee
which is invariant with respect to total Legendre transformations. Notice that the explicit components of the thermodynamic metric $g$
can be calculated in a straightforward manner once the fundamental equation $\Phi(E^a)$ is explicitly given.

%%%%%%%%%%%%%%%%%%%%%%%%%%%%%%%%%%%%%%%%%%%%%%%%%%%%%%%%%%%%%%%%%%%%%%%%%%%%%%%%%%%%%
%%%%%%%%%%%%%%%%%%%%%%%%%%%%%%%%%%%%%%%%%%%%%%%%%%%%%%%%%%%%%%%%%%%%%%%%%%%%%%%%%%%%%
\section{Quasi-static thermodynamic processes}
\label{sec:qua}

In ordinary thermodynamics, a quasi-static process is a thermodynamic process that happens infinitely slowly so that it can be ensured that 
the system will pass through a sequence of states that are infinitesimally close to equilibrium and, consequently, the system remains 
in quasi-static equilibrium. Since each point of the manifold ${\cal E}$ represents an equilibrium state, a quasi-static process 
can be interpreted as a sequence of points, i. e., as a curve in ${\cal E}$. In particular, the geodesic curves of ${\cal E}$ can represent 
quasi-static processes under certain conditions. 
A geodesic curve can be interpreted as a harmonic map from a 1-dimensional base space to the equilibrium manifold 
$({\cal E},g)$. The corresponding action represents a distance in ${\cal E}$ that we denote as the thermodynamic length $S=\int ds$ with 
$ds^2 = g_{ab}dE^a dE^b$. Then, the variation of the thermodynamic length leads to the geodesic equation
\be
\frac{d ^2E^a}{d\tau^2} + 
\Gamma^a_{\ bc} \frac{dE^b}{d\tau} \frac{dE^c}{d\tau} = 0 \ ,
\label{geo1}
\ee
where $\Gamma^a_{\ bc}$ are the Christoffel symbols of the thermodynamic metric $g$, and $\tau$ is an arbitrary affine parameter
along the geodesic. 

One can expect that  not all the solutions of the geodesic equations must be physically realistic.  
Indeed, there could be geodesic curves 
connecting equilibrium states that are not compatible with the laws of thermodynamics.
In particular, one would expect that the second law of thermodynamics imposes strong requirements on the solutions.  
In ordinary thermodynamics two equilibrium states are related to each other only if they 
can be connected by means of quasi-static process. Then, a geodesic that connects two physically
meaningful states can be interpreted as representing a quasi-static process. 
Since a geodesic curve is a dense succession of points, we conclude
that a quasi-static process can be seen as a dense succession of equilibrium states, a statement
which coincides with the definition of quasi-static processes in equilibrium thermodynamics  \cite{callen}.
Furthermore, the affine parameter $\tau$ can be used to label all equilibrium states which belong
to a geodesic. Since the affine parameter is defined up to a linear transformation, 
it should be  possible to choose it  in such a way that it increases as the 
entropy of a quasi-static process increases. This opens the possibility of interpreting the
affine parameter as a ``time" parameter with a specific direction which coincides with the
direction of entropy increase.

%%%%%%%%%%%%%%%%%%%%%%%%%%%%%%%%%%%%%%%%%%%%%%%%%%%%%%%%%%%%%%%%%%%%%%%%%%%%%%%%%%%%%%%
%%%%%%%%%%%%%%%%%%%%%%%%%%%%%%%%%%%%%%%%%%%%%%%%%%%%%%%%%%%%%%%%%%%%%%%%%%%%%%%%%%%%%%%
\section{Ordinary Thermodynamic systems}
\label{sec:gases}

The mathematical tools presented in the last sections allow us to define geometric structures 
in an invariant way. In particular, the curvature of the thermodynamic metric $g$ should 
represent the thermodynamic interaction independently of the thermodynamic potential. In fact,
this is not a trivial condition from a geometric point of view. For instance, a geometric analysis 
of black hole thermodynamics by using metrics introduced {\it ad hoc} in the equilibrium manifold
leads to contradictory results \cite{am,ama,aman06a,scws,caicho99,sst,med,mz,hernando2}. 
Using the induced thermodynamic metric
$g$ as defined in Section \ref{sec:eqm} for systems with second order phase transitions, the results are
consistent and invariant. To illustrate the formalism of GTD we now investigate the geometric 
representation of some ordinary thermodynamic systems.

%%%%%%%%%%%%%%%%%%%%%%%%%%%%%%%%%%%%%%%%%%%%%%%%%%%%%%%%%%%%%%%%%%%%%%%%%%%%
\subsection{The ideal gas}
\label{sec:ig}

As a concrete example of the application of GTD, we consider a mono-component 
ideal gas. This corresponds to the particular case $n=2$ of the metrics given 
in the last section. The corresponding fundamental equation can be written 
as $U(S,V)= [\exp(S/\kappa)/V]^{2/3}$, where $\kappa$ is a constant.
In this particular case, 
it turns out that the entropy representation is more convenient for the 
investigation of the field equations. To transform 
the results of the previous sections into the entropy representation, we notice 
that in this case the first law of thermodynamics is written as
$ dS= (1/T) dU + (P/T) d V$ 
so that the fundamental equation must be  given as $S=S(U,V)$, and the conditions
of thermodynamic equilibrium are $1/T = \partial S/\partial U$ and $P/T = 
\partial S /\partial V$. Consequently, in the entropy representation, 
the 5-dimensional 
phase manifold can be described by means of the coordinates 
\be
\label{coo}
Z^A = \left\{S,U,V,\frac{1}{T},\frac{P}{T}\right\}
\ee
and
the Riemannian metric (\ref{ginv2}) takes the form
\be
\label{gups}
G = \left(dS -\frac{1}{T} d U - \frac{P}{T} dV\right)^2
+ \Lambda \left[\left(\frac{U}{T}\right)^{2k+1} dU d\left(\frac{1}{T}\right) 
+\left(\frac{VP}{T}\right)^{2k+1} dV d\left(\frac{P}{T}\right)\right] \ .
\ee
Moreover, the explicit form of the 
Riemannian metric for the  equilibrium manifold can be derived
from Eq.(\ref{gdown1}). Then 
\bea
\label{gdowns}
g= \Lambda &\Bigg\{&\left(U\frac{\partial S}{\partial U}\right)^{2k+1}\frac{\partial^2 S}
{\partial U^2} dU^2
+  \left(V\frac{\partial S}{\partial V}\right)^{2k+1}
\frac{\partial^2 S}{\partial V^2} dV^2 \nonumber\\
& +&\left[ \left(U\frac{\partial S}{\partial U}\right)^{2k+1}
+ \left(V\frac{\partial S}{\partial V}\right)^{2k+1} \right] 
\frac{\partial^2 S}{\partial U \partial V} dU dV \ \Bigg\} \ .
\eea

It should be mentioned that this form of the thermodynamic metric is valid for any 
thermodynamic system with two degrees of freedom represented by the extensive
variables $U$ and $V$. It is only necessary to specify the fundamental equation 
$S=S(U,V)$ in order to completely determine the form of the metric.
In the specific case of an ideal gas, the fundamental equation can be expressed
as
\be
S(U,V) = \frac{3\kappa}{2}\ln U + \kappa\ln V\ .
\ee
A straightforward computation leads to the metric
\be
\label{gdownig}
g= - \kappa^{2k+2}\Lambda\left[ \left(\frac{3}{2}\right)^{2k+2} \frac{dU^2}{U^2}
+ \frac{dV^2}{V^2}\right]\ .
\ee
All the geometrothermodynamical information about the ideal gas must be contained 
in the metric (\ref{gdownig}). First, we must show that 
the subspace of equilibrium states $({\cal E},g)$ determines and extremal hypersurface
in the phase manifold $({\cal T}, G)$. 
The identification of the coordinates in ${\cal T}$ is as given in 
Eq.(\ref{coo}) so that the Christoffel symbols $\Gamma^A_{\ BC}$ for the metric
components $G_{AB}$ can be computed in a straightforward way. 
Then, the field equations can be reduced to 
\bea
\label{condig1}
\frac{\partial\Lambda}{\partial U} 
+ \frac{3\kappa}{2U^2} \frac{\partial \Lambda}{\partial Z^3}
+ 2(k+1)\frac{\Lambda}{U} & = & 0\ , \\
\frac{\partial\Lambda}{\partial V}
+ \frac{\kappa}{V^2} \frac{\partial \Lambda}{\partial Z^4}
+ 2(k+1)\frac{\Lambda}{V} & = & 0\ .
\label{condig2}
\eea
These are the conditions for the space of equilibrium states of the ideal gas to be
an extremal hypersurface of the thermodynamic phase space. Clearly, the arbitrariness
contained in the conformal factor $\Lambda$ allows us to find many solutions to the 
above equation. For instance, if we choose $\Lambda=$ const.  and $k=-1$, we obtain 
a particular solution which is probably the simplest one. This shows that the geometry
of the ideal gas is a solution to the motion equations of GTD. This special choice 
leads to the metric 
\be 
g=\frac{dU^2}{U^2} + \frac{dV^2}{V^2}
\label{met:ig}
\ee
whose curvature scalar vanishes identically. 
 This result agrees with our intuitive expectation that 
a thermodynamic metric with zero curvature should describe a system in which no
thermodynamic interaction is present. 

To continue the analysis of the geometry of the ideal gas we now investigate 
the geodesic equations. By means of the transformation $\xi= \ln U ,\eta= \ln V $,
the metric (\ref{met:ig}) takes the form  
\be
g = d\xi^2 + d\eta^2\ ,
\ee
where for simplicity  we set the additive constants of integration such that $\xi,\eta\geq 0$. 
The solutions of the 
geodesic equations are then found as $\xi = \xi_1 \lambda + \xi_0$
and   $\eta = \eta_1 \lambda + \eta_0$, where $\xi_0,\ \xi_1, \ \eta_0$ and $\eta_1$
are constants. This solution represents straight lines which on a $\xi\eta-$plane can be 
depicted by using the equation $\xi=c_1\eta + c_0$, with constants $c_0$ and $c_1$.
With our choice of integration constants, the only allowed range of values for $\xi$ 
and $\eta$ is within the quadrant determined by  $\xi\geq 0 $ and $\eta \geq 0$. 

\begin{figure}
\includegraphics[width=7cm]{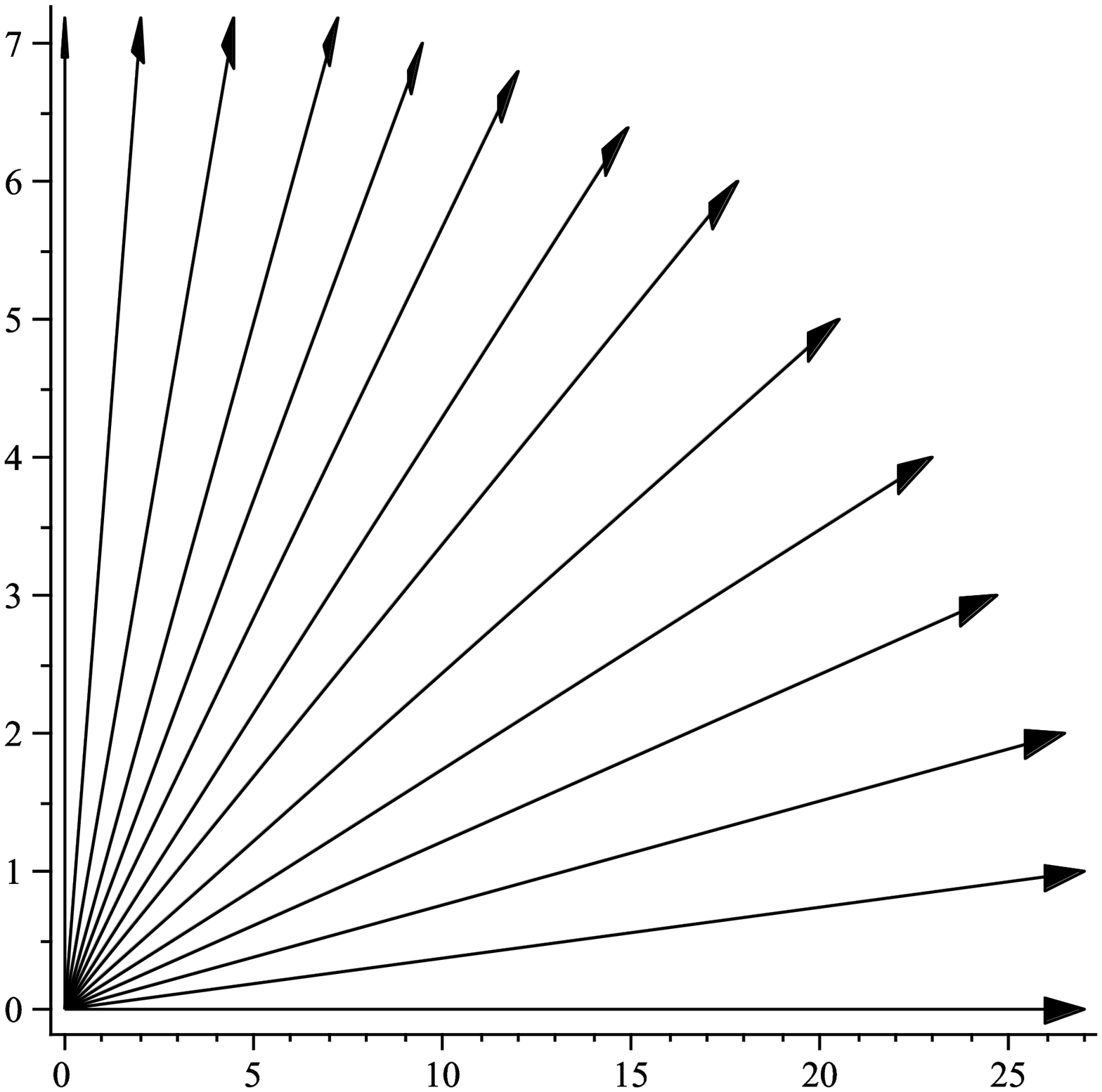}
\includegraphics[width=7cm]{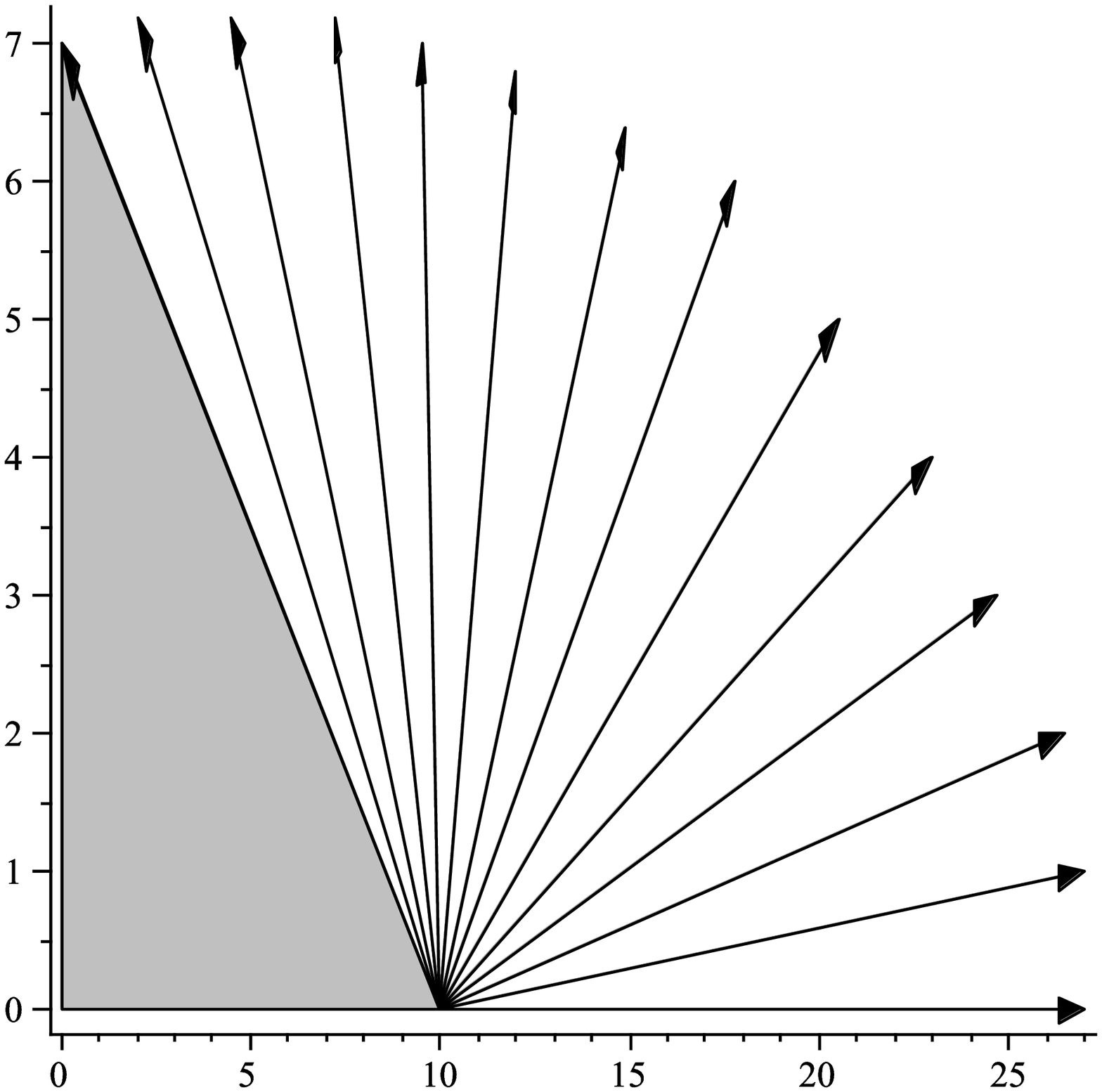}
\caption{Left figure: Geodesics in the space of equilibrium states of the ideal gas. All the states 
contained in the quadrant can be reached by only one geodesic which starts from
the initial state that coincides with the origin of coordinates. 
The arrows show the direction in which entropy increases.
Right figure: Geodesics with an initial state situated outside the origin of coordinates.
The shadow region contains all the states that due to the second law 
cannot be reached by geodesics with the fixed initial state. In all the geodesics
the ``arrow of time" is a consequence of the second law.}
\label{fig1}
\end{figure}

In this representation, the entropy becomes a simple linear function of the coordinates
and can 
be expressed as $S=(3\kappa/2)\xi + \kappa \eta$. Since each point on the
$\xi\eta-$plane can represent an equilibrium state, the geodesics should connect
those states which are allowed by the laws of thermodynamics. For instance, 
consider all geodesics with initial state $\xi=0$ and $\eta=0$. Then, any straight 
line pointing outwards of the initial zero point and contained inside 
the allowed positive quadrant connect states with increasing entropy. 
This behavior is schematically 
depicted in Fig.\ref{fig1} where the arrows indicate the direction 
in which a quasi-static process can take place. A quasi-static process connecting
states in the inverse direction is not allowed by the second law of thermodynamics. 
Consequently, the affine parameter $\tau$ along the geodesics can actually be 
interpreted as a time parameter and the direction of the geodesics indicates 
the ``arrow of time". If the initial state is not at the origin of the $\xi\eta-$plane,
the second law permits the existence of geodesics for which one of the coordinates,
say $\eta$, can decrease as long as the other coordinate $\xi$ increases in such a
way that the entropy increases or remains constant. This is 
schematically depicted in Fig.\ref{fig1} which also contains the region that cannot be reached
by geodesics. 

%%%%%%%%%%%%%%%%%%%%%%%%%%%%%%%%%%%%%%%%%%%%%%%%%%%%%%%%
\subsection{The van der Waals gas}
\label{sec:vdw}

A more realistic model of a gas, which takes into account the size of the particles 
and a pairwise attractive force between the particles of the gas, is based upon 
the van der Waals fundamental equation 
\be
\label{feqvdw}
S = \frac{3 \kappa}{2} \ln\left(U+ \frac{a}{V}\right) + \kappa \ln(V - b) \ ,
\ee
where $a$ and $b$ are constants. Usually, $a$ is interpreted as being 
responsible for the thermodynamic interaction, whereas $b$ plays a more
qualitative role in the description of the interaction \cite{callen}. 

The Riemannian structure of the manifold ${\cal T}$ is as before 
determined by the metric (\ref{ginv2}). For the sake of simplicity, 
we limit ourselves to the case with $k=-1$. Then, 
introducing the fundamental equation (\ref{feqvdw}) into the metric 
(\ref{gdown1}) with $k=-1$, the Riemannian structure of the manifold ${\cal E}$
is described by the metric 
\bea
\label{gdownvdw}
g = \frac{\Lambda}{U(U+a/V)} \Bigg[ && - dU^2 + \frac{U}{V^3} 
\frac{a(a+2UV)(3b^2-6bV+V^2)-2U^2V^4}{(V-b)(3ab-aV+2UV^2)}\, dV^2 \nonumber \\
&&+ \frac{a}{V^2}\frac{3ab-aV-3bUV+5UV^2}{3ab-aV+2UV^2} dU dV\Bigg] \ .
\eea
The curvature of this thermodynamic metric is in general non-zero, 
reflecting the fact that the thermodynamic interaction of the 
van der Waals gas is non-trivial. 
Furthermore, the scalar curvature  of the above metric can be written in the form
\be
R= \frac{a {\cal N}^{vdW}}{\left(P  V ^3 - a V + 2ab \right)^2 }
\ee
where ${\cal N}^{vdW}$  is a function of $U$, and $ V$
that is well--behaved at the points where the denominator vanishes. We see that
the scalar curvature  diverges at the critical points determined by the algebraic equation 
$P  V ^3 - a V + 2ab=0$. This is exactly the equation that determines the location of
first order phase transitions of the van der Waals gas \cite{callen}.  
Consequently, a first order phase transition
can be interpreted geometrically as a curvature singularity. This is in accordance with our intuitive interpretation of thermodynamic
curvature.

The motion equations (\ref{meng}) can be derived explicitly for this case
by using the phase manifold metric (\ref{ginv2}), with $k=-1$, and the metric 
(\ref{gdown1}) for the equilibrium manifold. 
It turns out that the motions 
equations reduce to only two first order partial differential equations
that can be expressed as
\bea
 \label{condvdw1}
\frac{\partial\Lambda}{\partial U} 
+ F_3 \frac{\partial \Lambda}{\partial Z^3}
+ F_4 \frac{\partial \Lambda}{\partial Z^4}
+ F_0 \Lambda & = & 0\ , \\
\frac{\partial\Lambda}{\partial V}
+ G_3 \frac{\partial \Lambda}{\partial Z^3}
+ G_4 \frac{\partial \Lambda}{\partial Z^4}
+ G_0 \Lambda & = & 0\ , 
\label{condvdw2}
\eea
where $F_0, F_3, F_4,G_0,G_3$, and $G_4$ are fixed rational functions of $U$ and $V$. 
Because of the arbitrariness of the conformal factor $\Lambda$ it is possible to find
solutions to the above system of partial differential equations. 
We conclude that a family of non-flat 
thermodynamic metrics can be found that determines an extremal hypersurface 
in the phase space, and can be used to describe the geometry of
the van der Waals gas. 

\begin{figure}
%\begin{center}
\includegraphics[width=9cm]{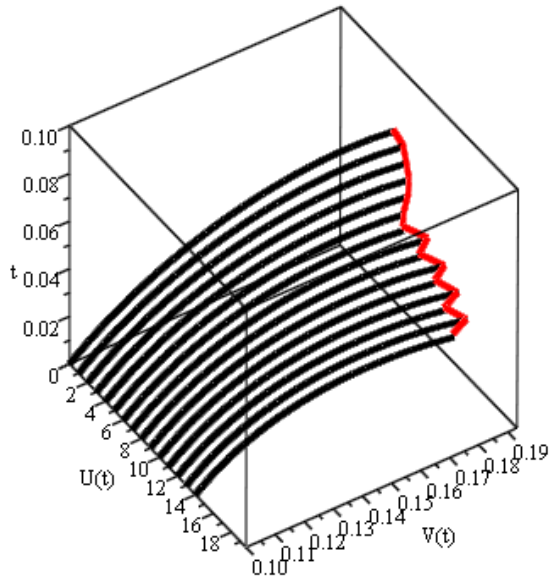}
\caption{Geodesics in the equilibrium manifold of the van der Waals gas for different initial values $U(\tau=0)$ and $V(\tau=0)=0.1$. The
end points of the thermodynamic variables are associated with first order phase transitions.  }
\label{fig2}
\end{figure}

The geodesic equations in the manifold described by the van der Waals 
metric (\ref{gdownvdw}) are highly non-trivial and require a 
numerical analysis \cite{tonio}. The results are illustrated in Fig.\ref{fig2}.
The main observation is that the geodesics are incomplete, i.e, there exist a maximum value 
of the affine parameter $\tau_{max}$ for which the numerical integration delivers an end 
value of $U(\tau_{max})$ and $V(\tau_{max})$. We analyzed numerically the end points 
 $U(\tau_{max})$ and $V(\tau_{max})$ and fount that at those points the relationship
$P  V ^3 - a V + 2ab=0$ is satisfied. We conclude that the geodesic incompleteness is due
to the appearance of first order phase transitions. Since geodesic incompleteness is 
usually associated with the existence of curvature singularities (see, for instance, \cite{hawellis})
the above result result corroborates the  fact that phase transitions correspond curvature singularities
in the equilibrium space.

%%%%%%%%%%%%%%%%%%%%%%%%%%%%%%%%%%%%%%%%%%%%%%%%%%%%%%%%%%%%%%%%%%%%%%%%%%%%%%%%%%%%%%%
%%%%%%%%%%%%%%%%%%%%%%%%%%%%%%%%%%%%%%%%%%%%%%%%%%%%%%%%%%%%%%%%%%%%%%%%%%%%%%%%%%%%%%%
\section{Conclusions}
\label{sec:con}

In this paper, we presented the most important mathematical elements of geometrothermodynamics (GTD), 
a formalism whose main goal is to describe in an invariant manner the properties of thermodynamic systems
by using geometric concepts. We use the concepts of contact geometry to define the thermodynamic phase 
manifold and to handle correctly the first law of thermodynamics and the Legendre transformations.
The phase manifold must be endowed with a Legendre invariant metric. We present the most general metric 
which is invariant with respect to partial and total Legendre transformations. If we limit ourselves to
the case of total Legendre transformations there are several metrics that preserve this symmetry. 
It turns out that it is necessary to use different metrics to describe thermodynamic systems with
either first order or second order phase transitions. We expect to explore in the near feature the 
cause of this difference. 

The equilibrium manifold is defined by means of a harmonic map in which the target space is the 
phase manifold. In this context, the equilibrium manifold turns out to be an extreme submanifold of the phase 
manifold endowed with a Riemannian thermodynamic metric which is determined uniquely in terms of the Legendre invariant 
metric introduced {\it ad hoc} in the phase manifold. The construction is such that only the fundamental equation
of the thermodynamic system is necessary in order to completely construct the geometry of the equilibrium 
manifold whose geometric properties are related to thermodynamic properties of the system. In particular, 
the thermodynamic interaction is described by means of the curvature, and phase transitions of the thermodynamic
system correspond to true curvature singularities of the equilibrium manifold. In this work, it was shown explicitly
that the curvature is a measure of the thermodynamic interaction in the case of the ideal gas and the van der Waals gas.
This statement has been confirmed in all the cases in which GTD has been applied so far 
\cite{aqs08,qs08,qs09a,qs09b,jjk10,mexpak10,chen11}.

As concrete examples of the application of GTD, we present the thermodynamic metric of the ideal gas and the van der Waals gas.
In the case of the ideal gas, the metric is flat as a result of the lack of thermodynamic interaction. In a particular 
coordinate system, the geodesics are represented as straight lines. Those geodesics which are in accordance with the laws
of thermodynamics turn out to represent quasi--static processes. In the case of the van der Waals gas, the metric 
is curved, indicating the presence of mechanical thermodynamic interaction between the constituents of the gas. 
True curvature singularities are found at those points where the gas undergoes a first order phase transition.
The geodesics of the equilibrium manifold of the van der Waals gas are shown to be incomplete at those points where
phase transitions occur. This could be used as an alternative method to find critical points where phase transitions
take place and curvature singularities exist.

%%%%%%%%%%%%%%%%%%%%%%%%%%%%%%%%%%%%%%%%%%%%%%%%%%%%%%%%%%%%%%%%%%%%%%
\section*{Acknowledgements}

This work was partially supported by DGPA-UNAM, grant No. 106110.

%%%%%%%%%%%%%%%%%%%%%%%%%%%%%%%%%%%%%%%%%%%%%%%%%%%%%%%%%%%%%%%%%%%%%%%%%%%%%%%%%%%%%%%


\begin{thebibliography}{99}


\bibitem{frankel} T. Frankel, {\em The Geometry of Physics: An Introduction} (Cambridge University Press, Cambridge, UK, 1997).

\bibitem{ym53} C. N. Yang and R. L.  Mills, Phys. Rev. {\bf 96}, 191 (1954).


\bibitem{greiner} W. Greiner, L. Neise and H. St\"ocker, {\it Thermodynamics and Statistical Mechanics} (Springer Verlag, New York, 
1995).


\bibitem{huang} K. Huang, {\it Statistical Mechanics} (John Wiley \& Sons, Inc., New York, 1987).

\bibitem{callen} H. B. Callen, {\it Thermodynamics and an Introduction to 
Thermostatics} (John Wiley \& Sons, Inc., New York, 1985).

\bibitem{rao45} C. R. Rao, Bull. Calcutta Math. Soc. {\bf 37}, 81 (1945).

\bibitem{amari85} S. Amari, {\it Differential-Geometrical Methods in Statistics} (Springer-Verlag, Berlin, 1985).


\bibitem{wei75} F. Weinhold, J. Chem. Phys. {\bf 63}, 2479, 2484, 2488, 2496 (1975); {\bf 65}, 558  (1976).

\bibitem{rup79} G. Ruppeiner, 
Phys. Rev. A {\bf 20},  1608 (1979).


\bibitem{rup95} G. Ruppeiner, Rev. Mod. Phys. {\bf 67},  605 (1995); {\bf 68},  313 (1996).


\bibitem{am} J. E. \AA man, I. Bengtsson, and N. Pidokrajt, Gen. Rel. Grav.
\textbf{35} 1733 (2003).

\bibitem{ama} J. E. \AA man and N. Pidokrajt, Phys. Rev. D \textbf{73}, 024017
(2006).

\bibitem{aman06a} J. E.  \AA man and N. Pidokrajt, Gen. Rel. Grav.\textbf{38},
1305 (2006).


\bibitem{scws} J. Shen, R. G. Cai, B. Wang, and R. K. Su, [gr-qc/0512035].

\bibitem{caicho99} R. G. Cai and J. H. Cho, Phys. Rev. D \textbf{60}, 067502 (1999).



\bibitem{sst} T. Sarkar, G. Sengupta, and B. N. Tiwari, J. High Energy Phys.
\textbf{0611} 015 (2006).

\bibitem{med} A. J. M. Medved, Mod. Phys. Lett. A \textbf{23}, 2149 (2008).

\bibitem{mz} B. Mirza and M. Zamaninasab,  J. High Energy Phys., 0706:059 (2007).    

\bibitem{hernando2} H. Quevedo, Gen. Rel. Grav. {\bf 40}, 971 (2008).


\bibitem{quev07} H. Quevedo, J. Math. Phys.
{\bf 48}, 013506 (2007).

\bibitem{her73} R. Hermann, {\it Geometry, physics and systems} (Marcel
Dekker, New York, 1973).


\bibitem{handbook} F. Dillen and L. Verstraelen, {\it Handbook of Differential Geometry} (Elsevier B. V., Amsterdam, 2006).


\bibitem{arnold} V. I. Arnold, {\it Mathematical Methods of Classical Mechanics}
(Springer Verlag, New York, 1980).

\bibitem{qstv10a} H. Quevedo, A. S\'anchez, S. Taj, and A. V\'azquez, Gen. Rel. Grav. DOI: 10.1007/s10714-010-0996-2 (2010).

\bibitem{aqs08} J. L. \'Alvarez, H. Quevedo, and A. S\'anchez,  Phys. Rev. D {\bf 77}, 084004 (2008). 

\bibitem{vqs09} A. V\'azquez, H. Quevedo, and A. S\'anchez, J. Geom. Phys. {\bf 60}, 1942 (2010).

\bibitem{qsv09} H. Quevedo, A. S\'anchez and A. V\'azquez, arXiv:math-phys/0811.0222 (2009).


\bibitem{misner} C. W. Misner, Phys. Rev. D {\bf 18}, 4510 (1978). 





\bibitem{tonio} A. Ram\'\i rez, Diploma thesis, Universidad Nacional Aut\'onma de M\'exico (2011), unpublished.

\bibitem{hawellis} S. Hawking and G. Ellis, {\it The large scale structure of space-time} (Cambridge University Press, Cambridge, UK, 1973).
 


\bibitem{qs08} H. Quevedo and A. S\'anchez,  JHEP {\bf 09}, 034 (2008). 

\bibitem{qs09a} H. Quevedo and A. S\'anchez, 
Phys. Rev. D {\bf 79}, 024012 (2009). 

\bibitem{qs09b} H. Quevedo and A. S\'anchez, 
Phys. Rev. D. {\bf 79}, 087504 (2009).

\bibitem{mexpak10} M. Akbar, H. Quevedo, K. Saifullah, A. S\'anchez, and S. Taj, 
{\it Thermodynamic Geometry Of Charged Rotating BTZ Black Holes},
arXiv:1101.2722 
 
\bibitem{jjk10} W. Janke, D. A. Johnston, and R. Kenna, 
{\it Geometrothermodynamics of the Kehagias-Sfetsos black hole},
arXiv:1005.3392.

\bibitem{chen11} P. Chen,  {\it 
Thermodynamic Geometry of the Born-Infeld-anti-de Sitter black holes}, arXiv:1104.0546



\end{thebibliography}
\end{document}